\documentclass[pop,fleqn]{w-art}
\usepackage{times}
\usepackage{w-thm}
\usepackage{graphicx}

\theoremstyle{plain}

\theoremstyle{definition}

\usepackage[]{graphicx}

\newcommand{\drawsquare}[2]{\hbox{%
\rule{#2pt}{#1pt}\hskip-#2pt%  left vertical
\rule{#1pt}{#2pt}\hskip-#1pt%  lower horizontal
\rule[#1pt]{#1pt}{#2pt}}\rule[#1pt]{#2pt}{#2pt}\hskip-#2pt%  upper horizontal
\rule{#2pt}{#1pt}}% right vertical
\newcommand{\fund}{\raisebox{-.5pt}{\drawsquare{6.5}{0.4}}}%  fund
\newcommand{\antifund}{\overline{\fund}}

\newcommand{\be}{\begin{equation}}
\newcommand{\ee}{\end{equation}}
\newcommand{\ba}{\begin{array}}
\newcommand{\ea}{\end{array}}
\newcommand{\bea}{\begin{eqnarray}}
\newcommand{\eea}{\end{eqnarray}}

\newcommand{\ov}{\overline}
\def\IR{\relax{\rm I\kern-.18em R}}
\def\o#1{\overline{#1}}
\def\IP{\relax{\rm I\kern-.18em P}}

\def\inbar{\vrule height1.5ex width.4pt depth0pt}
\def\IC{\relax\,\hbox{$\inbar\kern-.3em{\rm C}$}}
\def\K3{{\bf K3}}

\def\ov{\overline}

\def\n2d{\cN_{V^*}^{\otimes 2}}

\def\IC{\mathbb{C}}

\def\IR{\mathbb{R}}

\def\IP{\mathbb{P}}

\def\cN{{\mathcal N}}

\def\ch{\mbox{ch}}

\begin{document}
%%    The information for the title page will be placed between
%%    \begin{document} and \maketitle. The order of most entries
%%    is determined by the class file and can not be changed by
%%    rearranging them. The maketitle command follows after the
%%    abstract.
%%
%%    Most of the following commands will be completed by the publisher.
%%
%%    The copyrightyear is defined in the .clo file as the first argument
%%    of the copyrightinfo command. If the copyrightyear differs from that
%%    value it might be adjusted by the following definition:
%%
%% \renewcommand{\copyrightyear}{2007}% uncomment to change the copyrightyear.
%%
%\DOIsuffix{theDOIsuffix} %%
%% issueinfo for the header line
%\Volume{55} \Month{01} \Year{2007} %%
%%    First and last pagenumber of the article. If the option
%%    'autolastpage' is set (default) the second argument may be left empty.
\pagespan{1}{} %%
%%    Dates will be filled in by the publisher. The 'reviseddate' and
%%    'dateposted' (Published online) entry may be left empty.
%\Receiveddate{XXXX} \Reviseddate{XXXX} \Accepteddate{XXXX}
%\Dateposted{XXXX} %%
%\keywords{List, of, comma, separated, keywords.}
%\subjclass[pacs]{04A25%%PACS-Numbers
%\qquad\parbox[t][2.2\baselineskip][t]{100mm}{%
%  \raggedright
%  (Please use PACS-codes from the enclosed list
%  (ASCII2006FullPACS.txt) or from www.aip.org/pacs)\vfill}}%

%% \pretitle{Editor's Choice}

%% We have a short and a long form for the title. The short form
%% (optional argument) goes into the running head.

\begin{flushright} \vspace{-2cm}
{\small UPR-1190-T}
\end{flushright}

\title[D-brane instanton effects]{D-brane instanton effects in Type II orientifolds: local and global issues}

%% Please do not enter footnotes or \inst{}-notes into the optional
%% argument of the author command. The optional argument will go into
%% the header.  If there is only one address the marker \inst{x} may be
%% omitted.

%% Information for the first author.
\author[M. Cveti{\v c}]{Mirjam Cveti{\v c}\inst{1}%
%  \footnote{Corresponding author\quad E-mail:~\textsf{x.y@xxx.yyy.zz},
%            Phone: +00\,999\,999\,999,
%            Fax: +00\,999\,999\,999}
}
\address[\inst{1}]{Department of Physics and Astronomy, University of Pennsylvania, Philadelphia, PA 19104-6396, USA}
%%
%%    Information for the second author
\author[R. Richter]{Robert Richter\inst{1}}

%%
%%    Information for the third author
\author[T. Weigand]{Timo Weigand\inst{1}}
%%
%%    \dedicatory{This is a dedicatory.}

% \copyrightholder{Acoustical Scociety of America}
%\copyrightyear  {2001}

\begin{abstract}

We review how D-brane instantons can generate open string couplings of stringy
hierarchy in the superpotential which violate global abelian
symmetries and are therefore perturbatively forbidden. 
We discuss the main ingredients of this mechanism, focussing for
concreteness on Euclidean $D2$-branes in Type IIA orientifold
compactifications. Special emphasis is put on a careful analysis of
instanton zero modes and a classification of situations leading to
superpotential or higher fermionic F-terms. This includes the
discussion of chiral and non-chiral instanton recombination, viewed
as a multi-instanton effect. As phenomenological applications we
discuss the generation of perturbatively forbidden Yukawa couplings
in SU(5) GUT models and Majorana masses for right-handed neutrinos.
Finally we analyse the mirror dual description of $D1$-instantons in
Type I compactifications with $D9$-branes and stable holomorphic
bundles. We present globally defined semi-realistic string vacua on an elliptically
fibered Calabi-Yau realising the non-perturbative generation of
Majorana masses.

\end{abstract}

\date{\today}

\maketitle

\noindent Summary of talks given at the MCTP workshop  {\emph{The Physics and Mathematics of G2 Compactifications}} (T.W.), at String Phenomenology'07 (M.C.),  PASCOS'07 (M.C.),
BW'2007  (M.C.) and  at the Munich  {\emph{Workshop on recent developments in string effective actions and D-instantons}} (T.W., R.R.). \\
To appear in Fort. Phys.

\section{Introduction}
During the last year there has been some progress towards a better
understanding of non-perturbative effects in supersymmetric
four-dimensional string compactifications on Calabi-Yau orientifolds.
As was realized in \cite{Blumenhagen:2006xt,Haack:2006cy,Ibanez:2006da,Florea:2006si}, D-brane instantons can induce couplings between open string fields which are perturbatively forbidden because they violate global $U(1)$ selection rules. These effects are intrinsically stringy in that they cannot be described by conventional gauge instantons. For Type IIA orientifolds with intersecting D6-branes the relevant class of instantons is given by  Euclidean
$D2$-brane instantons, short $E2$-instantons, wrapping special
Lagrangian three-cycles of the internal Calabi-Yau space \cite{Blumenhagen:2006xt,Ibanez:2006da}.

In general the gauge group $U(N_a)$, carried by $N_a$ coincident
D-branes, contains an anomalous $U(1)_a$ which becomes massive via the generalized
Green-Schwarz mechanism and survives as a global
perturbative symmetry. It is due to this $U(1)$ selection rule that
particular phenomenologically important couplings are absent in intersecting brane
worlds. These include Majorana masses for the right-handed neutrino,
$\mu$-terms for the MSSM Higgs sector or Yukawa-couplings of type
${\bf 10}\cdot{\bf 10}\cdot{\bf 5}_H$ in $SU(5)$-like GUT models. On
the other hand, in the dual strongly coupled description in terms of
M-theory compactified on $G_2$-manifolds the $U(1)_a$ decouples
completely. There is therefore no associated selection rule and no
obstruction for the above mentioned couplings to exist. The resolution to this puzzle
is given by $U(1)_a$ breaking non-perturbative terms in the type IIA
picture.

Indeed, from the axionic shift symmetries under the abelian
symmetries induced by the Chern-Simons couplings of the $N_a$
$D6_a$-branes one finds the $U(1)_a$ transformation of the instanton
\cite{Blumenhagen:2006xt,Ibanez:2006da} \bea
 e^{-S_{E2}}=\exp\left[ \frac{2\pi}{ \ell_s^3}
           \left( -\frac{1}{g_s} {\rm Vol}_{\Xi} + i \int_{\Xi} C^{(3)}
         \right) \right] \longrightarrow  e^{i\, Q_a(E2)\,\Lambda_a}  \,\, e^{-S_{E2}}
\eea with \bea \label{chargee}
          Q_a(E2)={N}_a\,\, \Xi\circ (\Pi_a - \Pi'_a).
\eea Thus the exponential suppression factor $e^{-S_{E2}}$
characteristic for instantonic couplings transforms under the global
$U(1)$ symmetries in such a way that the full coupling \bea W_{np} =
\prod_i \Phi_{a_i b_i}\, e^{-S_{E2}} \eea is invariant again. In
this way an instanton with the appropriate zero mode structure has
the potential to generate perturbatively forbidden couplings.
%\cite{Blumenhagen:2006xt, Ibanez:2006da, Cvetic:2007ku,
%Ibanez:2007rs, Antusch:2007jd, Blumenhagen:2007zk}.

Various applications of the associated effects in different branches
of the string landscape have recently appeared in
\cite{Abel:2006yk,Akerblom:2006hx,Bianchi:2007fx,Cvetic:2007ku,Argurio:2007qk,Argurio:2007vq,Bianchi:2007wy,Ibanez:2007rs,Akerblom:2007uc,Antusch:2007jd,Blumenhagen:2007zk,Aharony:2007pr,Blumenhagen:2007bn,Aharony:2007db,Billo:2007sw,Billo:2007py,Sinha:2007rg,Aganagic:2007py,Camara:2007dy,Cvetic:2007qj,Ibanez:2007tu,GarciaEtxebarria:2007zv,Petersson:2007sc,Blumenhagen:2007sm,Bianchi:2007rb},
with related earlier work including
\cite{Witten:1996bn,Ganor:1996pe,Billo:2002hm}.

In the first part of this article we review some of the technical
aspects in dealing with stringy D-brane instantons, based mainly on
\cite{Blumenhagen:2006xt,Cvetic:2007ku,Blumenhagen:2007bn}. In
section \ref{Zero_sec} we give a detailed account of the zero mode
structure of D-brane instantons in Type II orientifolds, focusing
for concreteness on $E2$-instantons in Type IIA. In section
\ref{F-terms} we systematically distinguish between various
situations which lead either to genuine superpotential contributions
or to higher fermionic F-terms as discussed originally by Beasley
and Witten in the context of heterotic worldsheet instantons. A
novelty for half-BPS instantons in Type II orientifolds (as opposed
to heterotic theories) is the appearance of four instead of two
Goldstone fermions. In section \ref{sec_Rec} we discuss how these
modes can be lifted by couplings in the effective instanton action
which effectively describes the configuration as a two-instanton
process in the upstairs geometry before orientifolding
\cite{Blumenhagen:2007bn}. Unless further zero modes such as certain
charged or deformation modes remain unlifted, this shows that the
presence of these Goldstinos does not necessarily obstruct the
generation of a superpotential. We then exemplify, in section
\ref{Pheno_sec}, some phenomenological applications of stringy
instantons by detailing the generation of the aforementioned Yukawa
couplings in the context of SU(5) GUT models
\cite{Blumenhagen:2007zk} and  the generation of Majorana mass terms
\cite{Blumenhagen:2006xt,Ibanez:2006da}. The construction of
globally consistent string vacua exhibiting these effects turns out
to be easier in the mirror dual language of Type I
compactifications, which are the subject of section \ref{sec_TypeI}.
We apply the general setting to the construction of a class of
semi-realistic GUT type vacua on elliptically fibered manifolds
where the instanton contribution to Majorana masses can indeed be
proven to be non-vanishing \cite{Cvetic:2007qj}.

\section{Instanton zero modes}
\label{Zero_sec}

The computation of D-brane instanton effects hinges upon a careful analysis of the instanton zero modes.
For definiteness we focus now on compactifications of Type IIA on Calabi-Yau orientifolds with intersecting D6-branes
(see
\cite{Uranga:2003pz,Kiritsis:2003mc,Lust:2004ks,Blumenhagen:2005mu,Blumenhagen:2006ci}
for reviews).
The relevant spacetime instantons are given by
$E2$-branes wrapping special Lagrangian three-cycles $\Xi$ in the
Calabi--Yau so that they are point-like in four-dimensional
spacetime. In section \ref{sec_TypeI} of this article we describe the mirror symmetric picture of $E1$-instantons in Type I compactifications.

One distinguishes between two kinds of instanton zero modes
corresponding to whether or not they are charged under the gauge
groups on the $D6$-branes. The uncharged zero modes arise from the
$E2$-$E2$ sector. They always comprise the universal four bosonic
Goldstone zero modes $x^{\mu}$ due to the breakdown of
four-dimensional Poincar{\'e} invariance. Generically, for
instantons away from the orientifold fixed plane, these come with
four fermionic zero modes $\theta^{\alpha}$ and $\overline
\tau^{\dot \alpha}$
\cite{Argurio:2007qk,Argurio:2007vq,Bianchi:2007wy,Ibanez:2007rs}.
This reflects the fact that the instanton breaks half of the eight
supercharges preserved by the Calabi-Yau manifold away from the
orientifold fixed plane. The $\theta^{\alpha}$ modes indicate the
breakdown of one half of the ${\cal N}=1$ supersymmetry preserved by
the orientifold while the $\overline \tau^{\dot \alpha}$ modes are
associated with the breakdown of one half of its orthogonal
complement inside the ${\cal N}=2$ supersymmetry algebra respected
by the internal Calabi-Yau. For
 instantons on top of an orientifold plane these universal zero
modes are subject to the orientifold projection. For effective
$O6^{-}$-planes\footnote{ In contrast to effective $O6^{-}$-planes
effective $O6^{+}$-planes admit no supersymmetric tadpole
cancellation in $D6$-brane models and are thus not of
phenomenological interest.}, which we always assume in the sequel,
the non-dynamical gauge group of $N$ coincident instantons is $O(N)$ and $ x^{\mu}, \theta^{\alpha}$
are symmetrized, while the anti-chiral modes $ {\ov
\theta}_{\dot{\alpha}} $ are anti-symmetrized. Thus for a single
$O(1)$ instanton  the zero modes surviving the orientifold action
are $ x^{\mu}$ and $ \theta^{\alpha}$.

Generically the $E2-E2$ sector exhibits $b_1(\Xi)$ complex bosonic
zero modes $c_I$ associated with special Lagranigan deformations of the $E2$-instanton. Away
from the orientifold plane, each of these is accompanied by one
chiral and one anti-chiral Weyl spinor, $\chi^{\alpha}_I$ and $\ov
\chi^{\dot \alpha}_I$. For an instanton wrapping an $\Omega {\ov
\sigma}$ invariant cycle  $c_I, {\ov c}_I$ and $\ov \chi^{\dot
\alpha}_I$ are symmetrized while $\chi^{\alpha}_I$ are
anti-symmetrized or vice versa, depending on the type of cycle the instanton
wraps. In the sequel we refer to them as deformations of the first and
 second kind, respectively. In the T-dual Type I picture they correspond to the
position and Wilson line moduli of an $E1$-instanton wrapping a holomorphic curve.

Furthermore there arise zero modes at non-trivial intersections of
the instanton $E2$ with its orientifold image $E2'$. Intersections
away from the orientifold give rise to a chiral
supermultiplet $(m, \mu^{\alpha})$ and its anti-chiral counterpart.
For $\Omega {\ov
\sigma}$ invariant intersections the states
are subject to the additional orientifold projection which symmetrizes the zero
modes $m,{\ov m}, {\ov \mu}^{\dot{\alpha}}$ and anti-symmetrizes
the chiral mode $\mu^{\alpha}$. Thus for a single $U(1)$-instanton
we get the multiplicities for the zero modes as shown in table
\ref{antizero}.

\begin{table}[h]
\centering
\begin{tabular}{|c|c|c|}
\hline
 zero mode & $(Q_E)_{Q_{ws}}$ &   Multiplicity \\
\hline \hline
 $m, \ov m$ & $(2)_1$ ,$(-2)_{-1}$ & ${1\over 2}\left(\Xi'\circ \Xi+\Pi_{{\rm O}6}
\circ \Xi\right)$  \\
 $\ov\mu^{\dot \alpha}$ & $(-2)_{1/2}$ & ${1\over 2}\left(\Xi'\circ \Xi+\Pi_{{\rm O}6}
\circ \Xi\right)$  \\
 $\mu^{\alpha}$ & $(2)_{-1/2}$ & ${1\over 2}\left(\Xi'\circ \Xi-\Pi_{{\rm O}6}
\circ \Xi\right)$  \\
\hline
\end{tabular}
\caption{Charged zero modes on $E2-E2'$ intersection. % \vspace{3mm}
\label{antizero} }
\end{table}

This completes our discussion of the uncharged zero mode sector. In addition, there can arise fermionic
zero modes from intersections of the instanton $\Xi$ with
$D6$-branes $\Pi_a$ \cite{Blumenhagen:2006xt,Ibanez:2006da,Florea:2006si}.
%If the instanton is parallel to  $\Pi_a$,
%there are also massless bosonic modes in this sector. The detailed
%quantisation of these charged zero modes, both for chiral and
%non-chiral intersections, is described in \cite{Cvetic:2007ku}.
Let us focus for brevity on chiral intersections. As pointed out in
\cite{Cvetic:2007ku} states in the $E2-D6$ sector are odd under the
GSO projection contrary to the GSO-even states in the $D6-D6$ brane
sector. In particular, a positive intersection $I_{\Xi a} > 0$ of
the instanton and a D6-brane wrapping the respective cycles $\Xi$
and $\Pi_a$ hosts a  single \emph{chiral} fermion (i.e. with
world-sheet charge $Q_{\rm ws}=-\frac{1}{2}$) in the bifundamental
representation $(-1_E,\fund_a)$. The strict chirality of the charged
fermions is essential for the existence of holomorphic couplings
between these modes and open string states in the moduli action and
will also play a key role in the present analysis. For a generic
instanton cycle $\Xi$ away from the orientifold, this gives rise to
the charged zero mode spectrum summarised in table \ref{tablezero}.
 \begin{table}[ht]
\centering
\begin{tabular}{|c|c|c|}
\hline
zero modes&  Reps$_{Q_{ws}}$ & number   \\
\hline \hline
$\lambda_{a,I}$ &  $(-1_E,\fund_a)_{-1/2}$   & $I=1,\dots, [\Xi\cap \Pi_a]^+$    \\
$\overline{\lambda}_{a,I}$ &  $(1_E,\antifund_a)_{-1/2}$  & $I=1,\dots, [\Xi\cap \Pi_a]^-$    \\
\hline
$\lambda_{a',I}$ &  $(-1_E,\antifund_a)_{-1/2}$ & $I=1,\dots, [\Xi\cap \Pi'_a]^+$    \\
$\overline{\lambda}_{a',I}$    &  $(1_E,\fund_a)_{-1/2}$   & $I=1,\dots,[\Xi\cap \Pi'_a]^-$    \\
\hline
\end{tabular}
\caption{Zero modes at chiral $E2$ -$D6$ intersections.
\label{tablezero} } % \vspace{3mm}
\end{table}
As a result, the instanton carries the charge
$Q_a(E2)={\cal N}_a\,\, \Xi\circ (\Pi_a - \Pi'_a)$
 under the gauge group $U(1)_a$ \cite{Blumenhagen:2006xt,Ibanez:2006da}.

\section{Instanton generated F-terms}
\label{F-terms}
%Generically an instanton exhibits all types of zero modes described in the
%previous section.
All bosonic and
fermionic zero modes appear in the instanton measure. The
instanton induced amplitude is only non-vanishing if
all fermionic zero modes can be soaked up. In this section we analyse the possible instanton
generated F-term contributions, taking into account systematically which of the above zero modes are present in each case.

\subsection{Rigid $O(1)$ instantons}

We start with an $O(1)$ instanton wrapping a rigid cycle in the
internal Calabi-Yau. It therefore contains only the
four universal bosonic zero modes $x^{\mu}$ and the two chiral
fermionic zero modes $\theta^{\alpha}$ in the uncharged sector. If
in addition all charged zero modes arising at intersections with
$D6$-branes can be soaked up via disk diagrams\footnote{Here and in the sequel the subscripts
denote the worldsheet $U(1)$-charges. Disks containing more than one matter superfields are likewise possible.}
%\bea
$<\lambda^{a}_{-1/2}
\Phi^{ab}_{1} {\ov\lambda}^{b}_{-1/2}>$,
%>\,\,\,,\
%\eea
where $\Phi_{ab}= \phi_{ab}+\theta \psi_{ab} $ denotes
the chiral superfield arising at the intersection of branes $a$ and
$b$, such an instanton induces superpotential terms of the form \bea
\label{chirsuper}
            W\simeq  \prod_{i=1}^M  \Phi_{a_i,b_i}\, e^{-S_{E2}}.
\eea
Note that in case the product $\prod_{i=1}^M  \Phi_{a_i,b_i}$
is not invariant under the massive $U(1)_i$ such a coupling is
perturbatively forbidden and the instanton induced F-term represents the
leading contribution rather than just a correction to the effective
action.

For a systematic treatment of the associated instanton calculus in this context we refer to \cite{Blumenhagen:2006xt} and for its concrete application in a toroidal orbifold to \cite{Cvetic:2007ku}.
Due to the peculiar $\sqrt{g_s}$-scaling of the vertex operators of the charged zero modes, only disk diagrams containing precisely two $\lambda$-modes can contribute to the superpotential, while one-loop amplitudes with one end on the instanton and one end on the $D6$-branes do not carry any instantonic charged zero modes. Furthermore, the classical instanton suppression $S_{E2}= \frac{2\pi}{ \ell_s^3}( -\frac{1}{g_s} {\rm Vol}_{\Xi} + i \int_{\Xi} C^{(3)})$ receives corrections from one-loop amplitudes with no operator insertions \cite{Blumenhagen:2006xt}. These yield \cite{Abel:2006yk,Akerblom:2006hx} the threshold corrections \cite{Lust:2003ky,Akerblom:2007np} to the hypothetical gauge coupling associated with the cycle $\Xi$ and ensure holomorphicity of the superpotential \cite{Akerblom:2007uc,Billo:2007py,Akerblom:2007nh}.

\subsection{Non-rigid $O(1)$ instantons: Superpotential and higher fermionic terms}

Let us turn to an $O(1)$ instanton which does not wrap a rigid cycle
and thus carries additional deformation modes.
While deformation modes of the second kind
 induce corrections to the gauge kinetic function \cite{Akerblom:2007uc} we
focus on deformation modes of the first type. Generically, they generate higher fermionic couplings for the
closed string fields \cite{Beasley:2005iu}.
Before discussing these couplings let us examine
under what circumstances such an instanton could contribute instead to
the superpotential \cite{Blumenhagen:2007bn}. For simplicity we assume that
$b_1(\Xi)=1$. The
orientifold action projects out the chiral modes  $ {
\chi}^{\alpha}$ and the extra surviving
non-charged zero modes are $c, {\ov c} $ and ${\ov
\chi}^{\dot{\alpha}}$. For superpotential contributions to exist it
must be possible to absorb the fermionic zero modes ${\ov
\chi}^{\dot{\alpha}}$ without generating higher fermionic or
derivative terms. A way to do this for matter field superpotential
contributions is via a five-point amplitude \bea <{\ov
\chi}^{\dot{\alpha}}_{1/2}\,{\ov
\chi}^{\dot{\alpha}}_{1/2}\,\lambda^{a}_{-1/2} \Phi^{ab}_{0}
{\ov\lambda}^b_{-1/2} \label{superpotential}>.\eea If this five-point function has a contact term and if the remaining integral over
the bosonic instanton moduli space does not vanish, a
contribution to the superpotential can be generated. Let us stress
that from a general ${\cal N}=2$ SCFT point of view, no obvious
selection rules forbid such an interaction term. Having said this,
one can easily convince oneself that e.g. for factorisable three-cycles
on toroidal orbifolds the amplitude vanishes due to violation of the
$U(1)$ worldsheet charge which has to be conserved for each of the
three tori separately.
This, however, need not be so for more general setups and has to be determined in each case. \\
As observed in \cite{Beasley:2005iu} in the S-dual heterotic context such an
instanton (and its anti-instanton) can also generate higher fermionic F-terms which are
encapsulated in interactions of the form \bea S = \int d^4 x \, d^2
\theta \, \, w_{\ov i \ov j}\, (\Phi) {\ov{\cal D}}^{\dot \alpha}
{\ov \Phi}^{\ov i}  {\ov{\cal D}}_{\dot \alpha} {\ov \Phi}^{\ov j} + h.c.
\label{higher fermionic} \eea for the simplest case that the
instanton moves in a one-dimensional moduli space. Note that
supersymmetry requires a holomorphic dependence of $w_{\ov i \ov
j}\, (\Phi)$ on the superfields $\Phi$.

First let us reproduce the result associated with a deformation of the closed string moduli space obtained by Beasley and Witten in \cite{Beasley:2005iu} in the present context \cite{Blumenhagen:2007bn}.
Thus we assume no additional charged instantonic zero modes.
Denoting by ${\cal T} = T + \theta^{\alpha} t_{\alpha} $ the ${\cal
N}=1$ chiral superfield associated with the K\"ahler moduli, we can
absorb the instanton modulini by pulling down from the moduli action
two copies of the schematic anti-holomorphic coupling $\ov\chi^{\dot
\alpha} \ov t_{\dot
 \alpha}$.
In general the open-closed amplitude $\langle \ov \chi^{\dot \alpha}
\ov t_{\dot \alpha} \rangle$ does not violate any obvious selection
rule of the ${\cal N}=(2,2)$ worldsheet theory and is therefore
expected to induce the above coupling.
%\footnote{In particular, the
%total U(1) worldsheet charge is conserved. Still there might be
%situations, such as factorizable 3-cycles on $(T^2)^3$, where some
%of the separate U(1) charges are violated by this coupling. For
%details see appendix \ref{closed_int}.}.
Similarly, the two $\theta$-modes can be soaked up by the
holomorphic coupling  $\theta^{\alpha} u_{\alpha}$ involving the
fermionic partners of the complex structure moduli encoded in the
superfield ${\cal U} = U + \theta^{\alpha} u_{\alpha} $. This
results in a four-fermion interaction of the schematic form
$e^{-S_{E2}} \, u u \, \ov t \ov t$. Note that the coupling of the
complex and K\"ahler structure modulini only to the universal and
reparametrisation zero modes, respectively, is a consequence of
$U(1)$ worldsheet charges of the  associated vertex operators.
The derivative superpartner to the above
four-fermi term is obtained by integrating out two copies
of the term $\theta
\sigma^{\mu} \ov \chi  \, \partial_{\mu} \ov T$ in the instanton effective action, whose presence follows from evaluating the amplitude $\langle \theta^{\alpha} \,\, \ov
{\chi}^{\dot \alpha} \,\, \ov T \rangle$.
This can be summarized in superspace notation by
writing \bea \label{BW_2} S = \int d^4 x \, d^2 \theta \, \,
e^{-{\cal U}(\Xi)} \,\, f_{\ov i, \ov j}\left(e^{-{\cal T}_i},
e^{-\Delta_i}\right) {\ov{\cal D}}^{\dot \alpha} {\ov {\cal T}}^{\ov
i}  {\ov{\cal D}}_{\dot \alpha} {\ov {\cal T}}^{\ov j}, \eea where
${\cal U}(\Xi)$ is associated with the specific combination of
complex structure moduli appearing in the classial instanton action
and the holomorphic function $f_{\ov i, \ov j}$ depends in general
on the K\"ahler and open string moduli of the $D6$-branes
$\Delta_i$. This is indeed of the form \eqref{higher fermionic}
proposed in \cite{Beasley:2005iu}.

In the presence of a suitable number of charged $\lambda$ zero-modes
there exist, in addition to these closed string couplings, terms
which generate higher fermi-couplings also for the matter fields.
The Chan-Paton factors and worldsheet selection rules only allow the
$\lambda$ modes to couple holomorphically to the chiral open string
superfields, as for the generation  of a superpotential
\eqref{superpotential}, such that the instanton induces an
interaction as in (\ref{BW_2}), but with $ e^{-{\cal U}(\Xi)}$
simply replaced by  $ e^{-{\cal U}(\Xi)} \, \prod_{a_i,b_i}
\Phi_{a_i,b_i}$ (and modified coupling $f_{\ov i, \ov j}$).

For configurations with the correct charged zero mode structure, the
action can also pick up derivative terms directly involving the open
string fields. For this to happen the instanton moduli action has to
contain couplings of the form
%\bea
$ \ov {\chi}_{1/2}  \lambda^a_{-1/2}\,\,
            \ov\psi^{ab}_{1/2}\,\, \ov \lambda^b_{-1/2}$,
 where the fermionic matter field $\ov\psi_{1/2}^{\dot\alpha}$
lives at the intersection $D6_a-D6_b$ and lies in the anti-chiral
superfield $\ov\Phi$.

%\begin{figure}[h]
%\begin{center}
% \includegraphics[width=0.8\textwidth]{highF.eps}
%\end{center} \caption{\small Absorbtion of $\theta$ and
%$\bar{\chi}$-modes leading to F-terms. }\label{fermifig}
%\end{figure}

Integrating out two copies of this interaction term brings down the fermion bilinear $\ov\psi_{1/2}\ov\psi_{1/2} $. %characteristic for the higher fermionic terms  described in      \cite{Beasley:2005iu}.
In addition, the two $\theta^{\alpha}$ modes  again pull down a
bilinear of {\emph {chiral}} fermions $u^{\alpha}$ or, in the
presence of more $\lambda$ modes, $\psi_{ab}^{\alpha}$, as in the
case of superpotential contributions. This induces again a
four-fermi coupling. Alternatively, we can absorb one pair of
$\theta^{\alpha} \ov\chi^{\dot\alpha}$ via a disk amplitude of the
form  $<\theta_{3/2} \,\, \ov {\chi}_{1/2} \,\,
\lambda^a_{-1/2}\,\, \ov \phi_{-1} \,\, \ov \lambda^b_{-1/2}>$.
After bringing the $\ov \phi_{-1}$ into the zero ghost picture this
clearly generates a derivative coupling of the form $ \theta
\sigma^{\mu} \ov \chi  \,\lambda^a\, \partial_{\mu} \ov \phi \, \ov
\lambda^b $. Integrating out two copies of this term yields the
derivative superpartner to the above four-fermi term.

\subsection{Rigid $U(1)$-instantons not intersecting the O-plane}

Finally let us discuss a rigid $U(1)$ instanton which does not
intersect its orientifold image. This instanton contains all
the universal zero modes $x^{\mu}, \theta^{\alpha}$ and ${\ov
\tau}^{\dot{\alpha}}$. Such an instanton cannot contribute to the
superpotential unless the ${\ov \tau}^{\dot{\alpha}}$
modes are lifted by additional effects. This is what can happen e.g. for certain $E3$-instantons in Type IIB orientifolds in the presence of supersymmetric 3-form flux together with gauge flux on the instanton \cite{Blumenhagen:2007bn}.
Having said this, for gauge instantons, i.e. if the instanton wraps a cycle parallel to one of the spacetime-filling D-branes, the ${\ov
\tau}^{\dot{\alpha}}$ act as Lagrange multipliers which enforce the fermionic ADHM constraints \cite{Billo:2002hm}. This effect was generalised to stringy instantons parallel to a single $U(1)$-brane in \cite{Aganagic:2007py,Petersson:2007sc}.

In absence of such effects the instanton can nevertheless generate higher fermionic F-terms \cite{Blumenhagen:2007bn}.
The difference to the F-terms generated by a non-rigid $O(1)$
instanton discussed previously is that now only the complex
structure moduli receive derivative corrections. Denote by $w$ and
$a$ the scalar and axionic parts of the scalar component $U= w - i {
a}$ of a complex structure superfield. Then evaluation of the
amplitudes $\langle \theta \, \ov w \, \ov\tau \rangle$ and $\langle
\theta \,\ov a \, \ov\tau \rangle$ gives rise to the terms
$\theta \,\sigma^{\mu}\, \bar{\tau} \,\,
\partial_{\mu}\, \ov w^{i}, \, \theta \,\sigma^{\mu}\,
\bar{\tau} \,\, \partial_{\mu}\, \ov a^{i}$,  in the moduli
action. The absence of analogous terms for the K\"ahler moduli is a
consequence of $U(1)$ worldsheet charge conservation. Integrating
out two copies thereof indeed generates a derivative coupling of the
form $ e^{-S_{E2}} \, \partial \ov U \partial \ov U$. Together with
their fermionic partners, the derivative F-terms can be summarized
by \bea \label{BW_3} S = \int d^4 x \, d^2 \theta \, \, e^{-{\cal
U}(\Xi)} \,\, f_{\ov i, \ov j}\left(e^{-{\cal T}_i}, e^{\Delta_i}
\right)\,\,  {\ov{\cal D}}^{\dot \alpha} {\ov {\cal U}}^{\ov i} \,
{\ov{\cal D}}_{\dot \alpha} {\ov {\cal U}}^{\ov j} + \, \, \, h.c.,
\eea where the complex conjugate part is due to the anti-instanton
contribution. In the presence of charged zero modes $\lambda$ these
F-term corrections for the complex structure moduli involve
appropriate powers of charged open string fields required to soak up
the $\lambda$ modes. This amounts to replacing $e^{-{\cal U}(\Xi)}
\rightarrow e^{-{\cal U}(\Xi)} \prod_i \Phi_{a_i,b_i}$.

To summarize we have shown that not only rigid but under certain
circumstances also non-rigid $O(1)$-instantons give rise to superpotential contributions. The latter generically
generate higher fermionic F-terms \`a la Beasley and Witten involving
open and closed string superfields. Rigid $U(1)$-instantons with no
intersections with its orientifold image induce derivative
corrections to complex structure moduli space.

\section{Instanton recombination}
\label{sec_Rec} So far we have only considered instantons not
intersecting their orientifold image, i.e those without uncharged
zero modes arising from the $E2-E2'$ sector. For more generic
instantons exhibiting such zero modes to give contributions to the
superpotential the fermionic zero modes arising at the intersections
of $E2$ and $E2'$ need to be lifted in addition to the ${\ov
\tau}^{\dot{\alpha}}$ modes \cite{Blumenhagen:2007bn}. A reason to expect that this is
possible is the following: For $D6$-branes it is known that under
certain circumstances a pair of $D6$-$D6'$ branes can recombine into
a new sLag D6-brane which obviously wraps an $\Omega\o\sigma$
invariant three-cycle.  If a similar story also applies to pairs of
$E2-E2'$ instantonic branes, the recombined objects would be
candidates for new $O(1)$-instantons contributing to the
superpotential. Consequently, also the disjoint sum of $E2$ and
$E2'$ prior to recombination should yield a superpotential
contribution.
%In this section we investigate whether the naive expectation that
%such recombined $O(1)$-instantons exist is actually correct.

It turns out that an analysis of the $E2-E2'$ and $\ov \tau$ - modes is only part of the story. Recall that generically an instanton intersects the present $D6$-branes, which
gives rise to the previously described charged fermionic zero modes
$\lambda$. From table \ref{tablezero} the overall $U(1)_E$ charge
of these charged zero modes can be read off, \bea \label{chargetot}
\sum_i Q_E(\lambda^i) =  - \sum_a N_a \,  \Xi \circ (\Pi_a + \Pi_{a'}) = - 4 \,\, \Xi
\circ \Pi_{O6} . \eea Here  we have used the tadpole cancellation
condition.
This shows that in a globally consistent model  the total $U(1)_E$
charge of all charged zero modes is proportional to the chiral
intersection between the instanton and the orientifold plane. For an
$\Omega\o\sigma$ invariant instanton this last quantity vanishes,
whereas for a generic $U(1)$ instanton it does not.

If $\Xi \circ \Pi_{O6}\ne 0$, there must be additional
$U(1)_E$-charged zero modes in order for the zero mode measure to be
$U(1)_E$ invariant. These are the ones arising at the $E2-E2'$
intersection and displayed in table \ref{antizero}.

\subsection{Recombination of chiral $E2-E2'$ instantons}

In the following we assume that the instanton intersects its
image exactly once on top of the orientifold,
\bea
\label{int_pattern1} \Xi' \circ \Xi = \Pi_{O6} \circ \Xi=1.
\eea
As follows from the global consistency condition \eqref{chargetot}, the instanton suffers from
an excess of additional four charged zero modes ${\ov \lambda}$. The $U(1)_E$ invariant zero mode measure reads
\footnote{Note the inverse scaling behaviour of the Grassmann
numbers.}
\bea \label{measure} \int d{\cal M}_I = \int d^4 x\,  d^2
\theta d^2\,  \ov \tau \,\,  dm \, \, d \ov m  \, \, \underbrace{d^2
\overline \mu^{\dot \alpha}}_{Q_E = 4} \, \, \underbrace{ \prod_a
d\lambda_a\, \prod_b d \ov\lambda_b}_{Q_E = -4}.
\eea
%In analogy to
%D6-branes \cite{Kachru:1999vj}we expect that a slight deformation of
%the complex structure moduli induces a non-vanishing
%Fayet-Iliopolous term on the $E2$-world-volume  leading to
%condensation of the tachyonic uncharged zero modes fields $m$ and
%$\ov m$. This E2-brane recombination process preserves the
%topological charge of the intersecting $D6-D6'$ branes and therefore
%yields a supersymmetric brane wrapping a three-cycle which is
%invariant under $\Omega\ov\sigma$.

We first show that the uncharged modes in the $E2-E2'$ sector and the extra universal $ \ov \tau$ can successfully be lifted upon taking into account the interaction of the instanton with its orientifold image \cite{Blumenhagen:2007bn}. The two crucial couplings in the instanton effective action are
\bea \label{rec-action} S_{E2}= ( 2 m \, \ov m - \xi ) ^2 +m \, \ov
\tau_{\dot \alpha} \,\ov \mu^{\dot \alpha},
%+ m \, \ov \theta \,\ov \mu
\eea where the Fayet-Iliopoulos $\xi$ depends on the complex structure
moduli. It is proportional to the angle modulo
2 between the cycle $\Xi$ and its image $\Xi'$ and vanishes for
supersymmetric configurations. For $\xi$ positive the bosonic modes
$m$ become tachyonic and thus condense.
This results in a new instanton wrapping a new cycle with homology
class equal to $[\Xi] + [\Xi']$.
The instanton computation is performed for $\xi =0$, for which the CFT description of the effective action is valid.
Integrating out two copies of the second term in (\ref{rec-action}) saturates the extra uncharged zero modes in the instanton measure. Upon performing the path integral
over the bosonic modes this results in
%According to the second term in
%\eqref{rec-action} condensing $m$-modes, giving $m$ a non-vanishing
%VEV, induces a mass term for $\ov\tau$ and $\ov\mu$ and after
%integrating out the zero modes one is left with
the measure \bea \label{measure chiral} \int d{\cal M}_I = \int d^4
x\, d^2 \theta \underbrace{ \prod_a d\lambda_a\, \prod_b d
\ov\lambda_b}_{Q_E = -4}. \eea Although this measure has the correct
uncharged zero mode structure to give rise to superpotential terms
and thus looks quite encouraging, it turns out that there is no way
to soak up the excess $ \ov \lambda_b$ modes. Therefore the whole
instanton amplitude vanishes and in contrast to the naive
expectation the recombined $E2-E2'$ instanton does not contribute to
the superpotential.

\subsection{Recombination of non-chiral $E2-E2'$ instantons}

In the previous section we have seen that once $E2-E2'$ interactions are taken into acoount, even genuine $U(1)$ instantons
effectively have the correct  uncharged zero mode
structure to give rise to the superpotential. However, the additional $\ov
\lambda$ modes required to ensure global consistency spoil
the game. Consequently, it may be more promising to
consider an instanton intersecting non-chirally with its orientifold
image such that no net excess of charged $\lambda$ modes is needed
to satisfy \eqref{chargetot}.

The simplest non-trivial case involves one vector-like pair of zero
modes, i.e. \bea \label{int_pattern4} [\Xi' \cap \Xi]^+=[\Xi' \cap
\Xi]^- = 1, \ \
 [ \Pi_{O6} \cap \Xi]^+ =[ \Pi_{O6} \cap \Xi]^-= 1.
\eea  It gives rise to the  zero modes shown in table
\ref{antizeronc} and thus to the measure \bea \label{measure
non-chiral1} \int d{\cal M}_I = \int d^4 x\,  d^2 \theta d^2\,  \ov
\tau \,\,  dm \, \, d \ov m   dn \, \, d \ov n \, \, d^2 \overline
\rho^{\dot \alpha}\,\, d^2 \overline \nu^{\dot \alpha} \, \,
\underbrace{ \prod_a d\lambda_a\, \prod_b d \ov\lambda_b}_{Q_E = 0}
.\eea

\begin{table}[h]
\centering
\begin{tabular}{|c|c|}
\hline
 zero mode & $(Q_E)_{Q_{ws}}$  \\
\hline \hline
 $m, \ov m$ & $(2)_1$ ,$(-2)_{-1}$ \\
 $\ov\rho^{\dot \alpha}$ & $(-2)_{1/2}$ \\
\hline
 $n, \ov n$ & $(-2)_1$ ,$(2)_{-1}$ \\
 $\ov\nu^{\dot \alpha}$ & $(2)_{1/2}$ \\
\hline
\end{tabular}
\caption{Charged zero modes on non-chiral $E2-E2'$ intersection with
$O6^-$ plane.% \vspace{3mm}
\label{antizeronc} }
\end{table}
Generically the instanton moduli action takes the form \bea  S_{E2}=
( 2 m \, \ov m - 2n \ov n )^2 + \ov \tau_{\dot \alpha}(m \ov
\rho^{\dot \alpha} - n \ov \nu^{\dot \alpha}).
%+ m \, \ov \theta \,\ov \mu
\eea Performing the integrals over the fermionic and bosonic zero modes one is left with the instanton measure
 \bea
\label{measure non-chiral1} \int d{\cal M}_I = \int d^4 x\,  d^2
\theta d^2\,  \ov \tau \,\,  dm \, \, d \ov m   d^2 {\overline
{\widetilde \mu}}^{\dot \alpha} \, \, \underbrace{ \prod_a
d\lambda_a\, \prod_b d \ov\lambda_b}_{Q_E = 0} ,\eea where $\ov
{\widetilde \mu}^{\dot \alpha}_{1/2}= \ov\rho + \ov \nu$. Note that
this recombined instanton has precisely the same zero mode structure
as an $O(1)$ instanton with one deformation mode discussed in section
\ref{F-terms}. There we showed that if particular couplings are
present it can contribute to the usual superpotential F-terms \cite{Blumenhagen:2007bn}.
Additionally it generates higher fermionic F-terms \`a la Beasley and
Witten involving open and closed string superfields.

In certain situations there may be additional quartic couplings in the instanton effective action which allow one to integrate out also the net deformation modes
$\ov
{\widetilde \mu}^{\dot \alpha}_{1/2}$
\cite{GarciaEtxebarria:2007zv}. In this case, the $E2-E2'$ pair contributes to the superpotential also without invoking one of the mechanisms discussed in in section
\ref{F-terms}. Whether or not these terms are present can be read off uniquely from the dimension of the moduli space of the deformed cycle upon crossing the line of marginal stability. E.g. for non-chiral instanton recombination on toroidal orbifolds they are absent, as can be verified by a direct CFT computation.

%This is the open string sector which is invariant under
%$\Omega\o\sigma$ and gets symmetrized or anti-symmetrized (see
%appendix \ref{orientifold}) . Taking into account that the sign of
%the orientifold projection changes from $Dp$-$Dp$ to
%$D(p-4)$-$D(p-4)$ sectors, for a single $U(1)$ instanton we get the
%zero modes shown in Table \ref{antizero}.

%\begin{table}[h]
%\centering
%\begin{tabular}{|c|c|c|}
%\hline
% zero mode & $(Q_E)_{Q_{ws}}$ &   Multiplicity \\
%\hline \hline
% $m, \ov m$ & $(2)_1$ ,$(-2)_{-1}$ & ${1\over 2}\left(\Xi'\circ \Xi+\Pi_{{\rm O}6}
%\circ \Xi\right)$  \\
% $\ov\mu^{\dot \alpha}$ & $(-2)_{1/2}$ & ${1\over 2}\left(\Xi'\circ \Xi+\Pi_{{\rm O}6}
%\circ \Xi\right)$  \\
%\circ \Xi\right)$  \\
%\hline
%\end{tabular}
%\caption{Charged zero modes on $E2-E2'$ intersection % \vspace{3mm}
%\label{antizero} }
%\end{table}

\section{Phenomenological applications}
\label{Pheno_sec}

In this section we discuss two examples of instanton generated
perturbatively forbidden superpotential terms in Type IIA
orientifolds. As we will argue, the virtue of the instanton sector
is not only to generate such terms in the first place. As a bonus
the exponential suppression of the resulting operators with respect
to the string scale allows for a natural generation of certain
hierarchies which otherwise appear to be rather adhoc from a purely
four-dimensional point of view. A most prominent example of such a
hierarchical coupling is given by intermediate Majorana masses.
Before analysing their generation by instanton effects we discuss an
example of D-brane instanton generated Yukawa couplings in SU(5) GUT
models.
%Such instantons can contribute
%to the holomorphic superpotential only
%if they preserve half of the ${\cal N}=1$ supersymmetry.
%This means that the instanton measure must contain a factor
%$d^4 x\, d^2\theta$.

\subsection{$SU(5)$ Yukawa couplings}
Grand Unified $SU(5)$-like
models based on intersecting $D6$-branes generically suffer from the absence of the important Yukawa
coupling ${\bf 10}\cdot{\bf 10}\cdot{\bf 5}_H$ and are therefore
ruled out from being considered realistic.

%Such models were first generally proposed in
%\cite{Antoniadis:2000en} and explicitly constructed for intersecting
%D6-branes in
%\cite{Blumenhagen:2001te,Ellis:2002ci,Chen:2005ab,Gmeiner:2006vb}.

To construct such a model one only needs two stacks $a$ and $b$ of
branes realising the gauge group $U(5)_a\times U(1)_b$. The
$U(5)_a$ splits into $SU(5)_a\times U(1)_a$, where the anomalous
$U(1)_a$ gets massive via the generalized Green-Schwarz mechanism
and appears as a global symmetry in the effective action. While
matter transforming as $\bf{10}$ under $SU(5)_a$ arises at
intersections of stack $a$ with its image $a'$ the matter fields
transforming as $\bf{\bar{5}}$ as well as Higgs fields $\bf{5}_H$
and $\bf{\bar{5}}_H$ are located at intersections of stack $a$ with
$b$ and $b'$. Assuming that after applying the generalized Green
Schwarz mechanism only the combination $U(1)_X=\frac{1}{4}
U_a-\frac{5}{4}U(1)_b$ remains massless one obtains a flipped
$SU(5)\times U(1)_X$ model. Table \ref{tablegut} displays the
abstract quiver structure of such a setup.

\begin{table}[ht]
\centering
\begin{tabular}{|c|c|c|c|}
\hline
sector & number &  $U(5)_a\times U(1)_b$ reps. & $U(1)_X$   \\
\hline \hline
$(a',a)$ &  $3$ & ${\bf 10}_{(2,0)}$   & ${1\over 2}$   \\
\hline
$(a,b)$ &  $3$ & $\ov{\bf 5}_{(-1,1)}$   & $-{3\over 2}$   \\
\hline
$(b',b)$ &  $3$ & ${\bf 1}_{(0,-2)}$ & ${5\over 2}$     \\
\hline
$(a',b)$ &  $1$ & ${\bf 5}^H_{(1,1)}+\ov{\bf 5}^H_{(-1,-1)}$ & $(-1) + (1)$  \\
\hline
\end{tabular}
\caption{GUT $SU(5)$ intersecting D6-brane model.
\label{tablegut} } % \vspace{3mm}
\end{table}

Looking at the $U(1)_{a,b}$ charges it is clear that perturbatively
the two Yukawa couplings $ \bf 10\, \ov{\bf 5} \, \ov{\bf 5}^H$  and
$\ov{\bf 5}\, {\bf 5^H} \, {\bf 1}$ are present. For flipped $SU(5)$
these gives masses to the heavy (u,c,t)-Quarks and the leptons.
However, the Yukawa couplings for the light (d,s,b)-quarks \bea
 {\bf 10}_{(2,0)}\, {\bf 10}_{(2,0)}\, {\bf 5}^H_{(1,1)}
\eea are not invariant under the two $U(1)$s.

\begin{figure}[h]
\centering
 \includegraphics[width=0.8\textwidth]{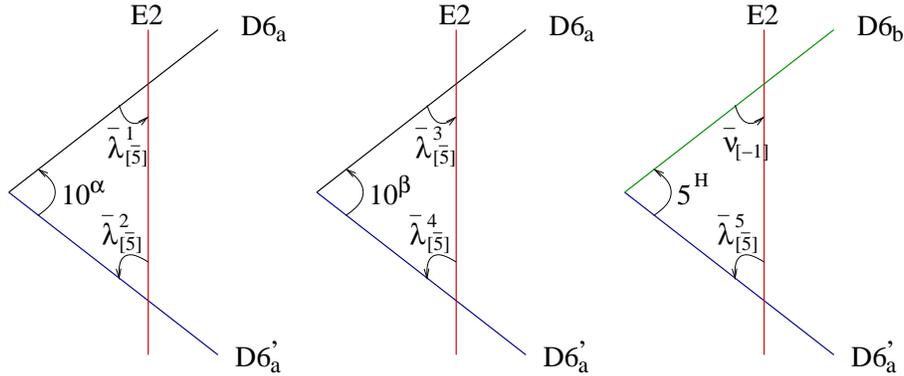}
\caption{\small Absorption of zero modes.}\label{yukafig}
\end{figure}

Now, if the model contains an $O(1)$-instanton with intersections
\cite{Blumenhagen:2007zk} \bea [\Xi\cap \Pi_a]^+=0, \quad [\Xi\cap
\Pi_a]^-=1, \quad [\Xi\cap \Pi_b]^+=0, \quad[\Xi\cap \Pi_b]^-=1,
\quad \eea then we get five zero modes $\ov\lambda^i_{[\ov 5]}$ from
the intersections of the instanton with $D6_a$ and one zero mode
$\ov\nu_{[-1]}$ from the intersection with $D6_b$. Since the
instanton lies in an $\Omega\ov\sigma$ invariant position, one can
absorb these six matter zero modes with the three disc diagrams
depicted in figure \ref{yukafig}.

All charge selection rules are satisfied. Due to the Grassmanian
nature of the fermionic zero modes $\ov\lambda^i$ the index
structure of the Yukawa coupling is \bea
    W_Y= x^{\alpha\beta}\, e^{-{2\pi\over \alpha_{GUT}}
  {\rm Vol_{E2}\over {\rm Vol}_{D6}} }\,\,\,
     \epsilon_{ijklm}\,\, {\bf 10}^\alpha_{ij}\, {\bf 10}^\beta_{kl}\, {\bf 5}^H_{m},
\eea where $\alpha,\beta=1,2,3$ are generation indices.
%The final instanton generate Yukawa coupling is of the form \bea
%              Y^{\alpha\beta}_{\langle{\bf 10}\, {\bf 10}\, {\bf 5}_H\rangle}=
%            x^{\alpha\beta}\, e^{-{2\pi\over \alpha_{GUT}}
%  {\rm Vol_{E2}\over {\rm Vol}_{D6}} }.
%\eea
 Here we have assumed that both D6-branes have the same volume
and gauge coupling unification at the GUT scale. The parameter
$x^{\alpha\beta}$ contains the contribution from the one-loop
determinants and additional prefactors of order one \cite{Cvetic:2007ku}.
%Note, that the one-loop
%contribution is by itself a sum over world-sheet instantons
%connecting the three intersection points in the disc diagrams like
%$10^\alpha\,\ov\lambda^i\ov\lambda^j$.
The location of the matter fields depend on the flavour index and
therefore also the Yukawa couplings do.

To summarize, we find that D-brane instantons can generate the ${\bf
10}\, {\bf 10}\, {\bf 5}_H$ Yukawa coupling. For the case of flipped
SU(5) the hierarchy between the $(u,c,t)$ quarks and the $(d,s,b)$
quarks is explained by the $E2$-instanton suppression
$\exp(-S_{E2}(U))$, whereas the hierarchy between the first, second
and third family is due to the different suppression of world-sheet
disc instantons $\exp(-S_{ws}(T))$, where $U$ denote the complex
structure and $T$ the K\"ahler superfields.

\subsection{Majorana masses}
\label{Majorana_sec1}
Another interesting perturbatively forbidden coupling is the
Majorana mass term for right-handed neutrinos. These Majorana masses
together with non-perturbative Dirac masses give rise to the seesaw
mechanism. Generically the right-handed neutrino is located at
intersections between two abelian gauge groups $U(1)_a$ and $U(1)_b$
which become massive due to the generalized Green-Schwarz mechanism. In
order to realise the remaining matter content of the MSSM we have to
introduce additional branes. For simplicity let us focus on a
supersymmetric $SU(5)$ model, where we need only one additional stack of 5 coincident
branes providing an $SU(5)$ gauge group. Table \ref{tablemajorana}
displays the matter
content as well as the associated representations under the gauge groups.\\

\begin{table}[ht]
\centering
\begin{tabular}{|c|c|c|c|}
\hline
sector & number &  $U(5)_a\times U(1)_b\times U(1)_c$    \\
\hline \hline $(a',a)$ &  $3$ & ${\bf 10}_{(0,0)}$\\
 \hline
$(a,c)$ &  $3$ & $\ov{\bf 5}_{(1,0)}$     \\
\hline
$(b,a)$ &  $1$ & ${\bf 5}^H_{(1,0)}+\ov{\bf 5}^H_{(-1,0)}$     \\
\hline
$(b,c)$ &  $3$ & ${\bf 1}_{(-1,1)}$   \\
\hline
\end{tabular}
\caption{GUT $SU(5)$ intersecting D6-brane model.
\label{tablemajorana} } % \vspace{3mm}
\end{table}

%\begin{table}[ht]
%\centering
%\begin{tabular}{|c|c|c|c|}
%\hline
%sector & number &  $U(5)\times U(1)_a\times U(1)_b$    \\
%\hline \hline
%$(c',c)$ &  $3$ & ${\bf 10}_{(0,0)}$     \\
%$(c,b)$ &  $3$ & $\ov{\bf 5}_{(0,-1)}$     \\
%$(a,c)$ &  $1$ & ${\bf 5}^H_{(1,0)}+\ov{\bf 5}^H_{(-1,0)}$     \\
%$(a,b)$ &  $3$ & ${\bf 1}_{(-1,1)}$   \\
%\hline
%\end{tabular}
%\caption{GUT $SU(5)$ intersecting D6-brane model
%\label{tablemajorana} } % \vspace{3mm}
%\end{table}
We note that the coupling generating the Dirac mass is realized
$\ov{\bf 5}\,{\bf 5}^H\,  {\bf 1}$ while a Majorana mass term for
the righthanded neutrino  \bea M_{N^{c}_{R}}\,{\bf 1}_{(-1,1)}\,{\bf
1}_{(-1,1)}\, \eea is forbidden due to the global $U(1)_b$ and
$U(1)_c$. If the model contains an $O(1)$-instanton with
intersections \cite{Blumenhagen:2006xt,Ibanez:2006da, Cvetic:2007ku,
Ibanez:2007rs,Antusch:2007jd} \bea [\Xi\cap \Pi_b]^+=0, \quad
[\Xi\cap \Pi_b]^-=2, \quad [\Xi\cap \Pi_c]^+=2, \quad[\Xi\cap
\Pi_c]^-=0, \quad \eea we get
%apart from the universal zero modes $x^{\mu}$, $\theta^{\alpha}$ four
%additional zero modes $\bar{\lambda}^{1,2}_{a}$ and
%$\lambda^{1,2}_{b}$,
the right number of charged zero modes which can be absorbed via two
disc diagrams.
%shown in figure \ref{majorana}.
%\begin{figure}[h]
%\centering
% \includegraphics[width=0.30\textwidth]{majorana.eps}
%\caption{\small Absorption of zero modes}\label{majorana}
%\end{figure}
Performing the integration over the instantonic zero modes yields
\cite{Blumenhagen:2006xt,Cvetic:2007ku}\bea M^{\alpha\,
\beta}_{N^{c}_{N_R}}= x^{\alpha\,\beta}\,M_s
 e^{-{2\pi\over \alpha_{GUT}}
  {\rm Vol_{E2}\over {\rm Vol}_{D6_a}} },
\eea where $M_s$ is the string mass and $\alpha_{GUT}$ denotes the
gauge coupling at the GUT scale. The prefactor $x$ receives
contributions from the one-loop determinant as well as from the disc
computation and is of order $1$. Again the result is flavor
dependent due to world sheet instantons. For a ratio of the volumes
of $0.06<{\rm Vol_{E2}\over {\rm Vol}_{D6_a}}<0.02 $ we obtain
Majorana masses in the range of $10^{11}-10^{15} GeV$ which yield, together
with the perturbative Dirac neutrino masses, the desired hierarchically small see-saw masses. For a
local supersymmetric realization as well as the computation of the
instanton induced Majorana mass term for right-handed neutrino see
\cite{Cvetic:2007ku}.

These examples illustrate the importance of taking into account
non-perturbative effects in exploring the string landscape. It
is particularly desirable to discover globally defined
realistic string vacua in which such and similar genuinely stringy
effects naturally realize the hierarchies of scale which seem so
puzzling from a purely four-dimensional effective field theoretical
point of view. As we will see in the next section, such vacua can indeed be constructed in the framework of Tpye I
compactifications.

\section{$E1$-instantons in Type I compactifications}
\label{sec_TypeI}

While the previous description of D-brane instantons is very illustrative in
Type IIA language, Type IIB orientifolds are amenable to
the techniques of complex geometry. This facilitates the
construction of globally defined models. For this purpose we now
consider the mirror symmetric formulation of stringy instantons in
Type I compactifications on an internal Calabi-Yau threefold $X$.
The gauge sector is defined in terms of stacks of $M_a=n_a \times
N_a$ spacetime-filling $D9$-branes wrapping the whole of $X$ (and
their orientifold images), where $\sum_a n_a N_a = 16$.  These
$D9$-branes can carry rank $n_a$ holomorphic vector bundles $V_a$
whose structure group $U(n_a)$ breaks the original gauge group
$U(M_a)$ associated with the coincident $D9$-branes to the commutant
$U(N_a)$ \cite{Blumenhagen:2005zg, Blumenhagen:2005zh}. Stacks of
$N_i$ $D5$-branes wrapping the holomorphic curve $\Gamma_i$ on $X$
carry gauge group $Sp(2N_i)$.

The massless open string spectrum is encoded in various cohomology
groups associated with the respective bundles on the branes \cite{Blumenhagen:2005zg, Blumenhagen:2005zh}.
Cancellation of $D5$-tadpoles requires $\sum_a N_a \ch_2(V_a) - \sum_i
N_i \gamma_i = - c_2(TX)$, and absence of global anomalies is
ensured by the constraint $\sum_a N_a c_1(V_a) \in H^2(X,
2{\mathbf Z})$.

%The vector bundles must be supersymmetric with respect to the O9-plane. For stable holomorphic
%bundles this amounts to satisfying the D-flatness condition inside the K\"ahler cone, $\int_X {1
%\over 2}     J\wedge J \wedge c_1(V_a)  -
% \ell_s^4 \,
%\left( {\rm ch}_3 (V_a)+\frac{1}{24}\, c_1(V_a)\,  c_2(T)\right)=0$, which can be read as a
%constraint on the K\"ahler form $J$. Finally the real part of the gauge kinetic function $Re(f_a)=
%{\widetilde f}_a/(2 \pi g_s \ell_s^6)$ has to be positive. Here \bea \label{Gauged} {\widetilde f}_a=
%{n\over 3!}\,
% \int_X J \wedge J \wedge J -
% \ell_s^4\,  \int_X J \wedge
%\left( {\rm ch}_2 (V_a)+\frac{n_a}{24}\, c_2(T)\right). \nonumber \eea

The superpotential of the four-dimensional ${\cal N}=1$
supersymmetric effective action receives non-perturbative
corrections due to $E1$-instantons on a holomorphic curve $C$
\cite{Witten:1999eg}. For simplicity we focus on $E1$-branes
wrapping rigid isolated ${\mathbf P}^1$s, which correspond on the Type IIA side to rigid $O(1)$-instantons. As described in the Type
IIA context, in the presence of D-branes charged zero modes
$\lambda$ arise, this time in the $D9-E1$ sector. In the conventions
of \cite{Cvetic:2007qj}, fermionic zero modes in the representation $({N}_a,1_E)$
are counted by the cohomology group $H^0(C, V^{\vee}_a|_C \otimes
K_C^{1/2})$, while the zero modes in the conjugate representation
$(\ov N_a,1_E)$ are associated with $H^0(C, V_a|_C \otimes
K_C^{1/2}) = H^1(C, V^{\vee}_a|_C \otimes K_C^{1/2})^*$, see table
\ref{lambda_cohom}.

\begin{table}[htb!]
\centering
\begin{tabular}{|c|c|c|}
\hline \hline
state & rep & cohomology    \\
\hline \hline
$\lambda_a$  & $(N_a,1_E)$   & $H^0({\mathbf P}^1, V_a^{\vee}(-1)|_{{\mathbf P}^1})$  \\
$\ov\lambda_a$  & $(\ov N_a,1_E)$   & $H^1({\mathbf P}^1, V_a^{\vee}(-1)|_{{\mathbf P}^1})^*$ \\
\hline \hline
\end{tabular}
\caption{\small Fermionic zero modes in $D9-E1$ sector. }
\label{lambda_cohom}
\end{table}

These zero modes are particularly well under control if the
$D9$-branes carry rank one, i.e. complex line bundles $L_a$. From the
above it follows, with $K_{{\mathbf P}^1} = {\cal O}(-2)$, that the
relevant object is $L_a(-1)|_{{{\mathbf P}^1}} = {\cal O}(x_a-1)$,
where $x_a= \int_{{\mathbf P}^1} L_a$. The charged zero modes are
then readily determined  by Bott's theorem. Additional fermionic
zero modes from the $D5-E1$ sector are
 counted by the extension groups $Ext_X(j_* {\cal
O}|_{\Gamma_i},i_* {\cal O}|_C)$. These groups vanish when
$\Gamma_i$ and $C$ do not intersect.

If all $\lambda$ modes can be absorbed consistently the $E1$-instanton
yields contributions to the superpotential of the schematic form
\bea \label{super} W=  M_s^{3-k} \prod_{i=1}^k \Phi_i e^{- \frac{2
\pi}{g_s} {{\mathcal Vol} \over \ell_s^2} } =  M_s^{3-k} \prod_i
\Phi_i  e^{-  \frac{2 \pi \, \ell_s^4 }{\alpha_{a}} \frac{{\mathcal
Vol}}{{\widetilde f_a}} }. \eea Here we traded $g_s$ in for the
gauge coupling on a reference brane $D9_a$. The  scale of the
non-perturbative term is thus controlled by the ratio of the
instanton volume ${\mathcal Vol}= \int_{{\mathbf P}^1} J$ to the
gauge kinetic function ${{\widetilde f_a}}$. This ratio is a
function of the K\"ahler moduli which are in general only partially
constrained by the D-flatness conditions. In (\ref{super}) we also
suppressed the possible dependence on the complex structure moduli
through the one-loop Pfaffian
\cite{Blumenhagen:2006xt,Akerblom:2006hx}.

\subsection{$E1$-instantons on elliptic CY3}

\begin{figure}[h]
\centering
 \includegraphics[width=0.55\textwidth]{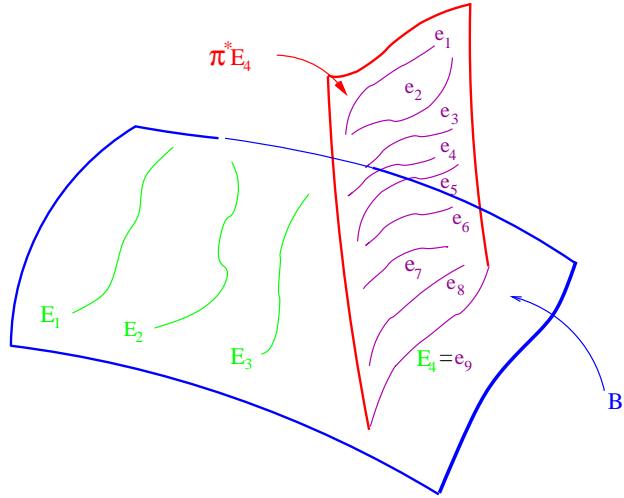}
\caption{\small The $dP_9$ surface $\pi^*E_4$ inside the firbration $\pi:X\rightarrow B.$}\label{dP9fig}
\end{figure}

\noindent We now present a class of  globally defined supersymmetric
models exhibiting such instanton effects. As Calabi-Yau threefold
$X$ we choose a generic elliptic fibration $\pi: X \rightarrow B$
over the del Pezzo surface $ B=dP_r$ for $r=4$.
%For detailed information on this
%type of geometries see, e.g., \cite{Donagi:1999gc}. For concrete
%algebraic geometry formulae relevant in our context see
%\cite{Blumenhagen:2005zg}.
The K\"ahler form $J$ of X enjoys the expansion $J/\ell_s^2=
r_{\sigma} \sigma + r_l \pi^* l + \sum_{i=1}^4 r_i \pi^* E_i$  in
terms of the fibre class $\sigma$ as well as the pullbacks of the
the hyperplane class $l$ and the classes $E_i$ of the four ${\mathbf
P}^1$s inside $dP_4$ obtained as the blow-up of certain
singularities in ${\mathbf C}{\mathbf P}^2$. The intersection form
on $B$ is $l\cdot l =1, E_i \cdot E_j = -\delta_{ij}$. An obvious
class of rigid isolated ${\mathbf P}^1$ in $X$ consists in the ten
horizontal ${\mathbf P}^1$s inherited from the base $B$, $E_i$ and
$l-E_i-E_j, i \neq j$. They are described as the intersection of the
zero section $\sigma$ with the divisors $\pi^*E_i$ or
$\pi^*(l-E_i-E_j)$. Each  $\pi^*E_i$ and $\pi^*(l-E_i-E_j)$ is
itself an elliptic fibration over the respective horizontal $\mathbf
P^1$  and thus represents a $dP_9$ surface. As such it  contains an
infinite number of rigid isolated ${\mathbf P}^1$s, which are also
rigid as curves in $X$. Consider for definiteness the divisor
$\pi^*E_4$. Its second homology class $H_2(\pi^*E_4, \mathbf Z)$ is
spanned by the hyperplane class $h$ and the classes of the nine
$\mathbf P^1$s $e_i, i=1, \ldots, 9$ with intersection form $h\cdot
h =1, e_i \cdot e_j = -\delta_{ij}$ (see figure \ref{dP9fig}).
However, these classes are not independent as elements in $H_2(X,
\mathbf Z)$, but are mapped under $\phi: H_2(\pi^*E_4, \mathbf Z)
\rightarrow H_2(X, \mathbf Z)$ to \bea \phi(e_i) = f + \sigma
\pi^*E_4, \quad i=1, \ldots 8, \quad \phi(e_9) = \sigma \pi^*E_4,
\quad \phi(h) = 3(f + \sigma \pi^*E_4), \eea where $f$ denotes the
fiber class as an element in $H_2(X, \mathbf Z)$. In particular, we
identify $e_9$, the base of the fibration $\pi^*E_4$, with the
horizontal curve $\sigma \pi^*E_4$. The set of rigid isolated
$\mathbf P^1$s in $\pi^*E_4$ is given by all integer combinations
$d_0 h + \sum_i d_i e_i$ subject to $ 3d_0 + \sum_i d_i = 1 =
\sum_i d_i^2 - d_0^2 $ \cite{Donagi:1996yf}. It follows that their
class in $X$ is given by $(f + \sigma \pi^*E_4) + d_9 \sigma
\pi^*E_4$.  In the sequel we will be interested in the effect of
$E1$-instantons wrapping one of these curves.

For our purposes it suffices to consider branes carrying line
bundles $L_a$ with first Chern class $c_1(L_a)= q_a \sigma +
\pi^*\zeta_a$ with $\zeta_a \in H^2(dP_4, {\mathbf Z})$. For the
computation of charged zero modes of an instanton wrapping one of
the above classes of curves it is important to note that
$L_a|_{{\mathbf P}^1}= {\cal O} (\zeta_a \cdot E_4 - q d_9)$. Note furthermore
that a D5-brane wrapping a ${\mathbf P}^1$ with $d_j =0$ does not
intersect the curve $e_j$ and therefore introduces no unwanted zero
modes on an instanton wrapping $e_j$. On the other hand, those
curves intersecting the D5-brane do carry additional zero modes and
therefore yield different contributions to the effective action.
This provides a way round the cancellation results of \cite{Beasley:2003fx} for
certain dual heterotic (0,2) non-linear $\sigma$-models on complete
intersection manifolds.

%For the concrete expressions of all higher Chern characters and the Euler characteristic
%$\chi(L_a)$ in terms of $c_1(L_a)$  see sections 3.3 and 3.4
%of \cite{Blumenhagen:2005zg},
%while the full cohomology groups $H^*(X, L_a)$ can be computed  from
%appendix B in \cite{Blumenhagen:2006wj}.

%We consider instantons wrapping certain rigid non-horizontal ${\mathbb P}^1$s
%in the $dP_9$ surface $\pi^*E_4$, which is obtained as the pullback of the ${\mathbb P}^1$ $E_4$ in the base.
%Specifically, the ${\mathbb P}^1$s are taken not to intersect the base of $dP_9$.
%This choice ensures
%that for the above type of line bundles one can always take each of the filler D5-branes to wrap a purely horizontal or vertical curve $\Gamma_i$ that
%does not hit the
%instanton, thus introducing no extra zero-modes.
%It turns out that one can always take the filler D5-branes to wrap curves $\Gamma_i$ that do not hit at least $\emph{some}$ of the ${\mathbb P}^1$s in that class, thus introducing no extra zero modes for this subset of instantons.

\subsection{Majorana masses}

The described framework allows for the realisation of instanton
generated superpotential couplings of various types in concrete,
globally defined string vacua. This was demonstrated in \cite{Cvetic:2007qj} in
the context of an SU(5) GUT model with instanton generated Majorana
masses of the type described in section \ref{Majorana_sec1}. In a second class of
GUT models presented in \cite{Cvetic:2007qj}, instantons generate a superpotential of Polonyi type in
the hidden sector as a step towards hierarchically suppressed
dynamical supersymmetry breaking.

%\noindent In models with instanton-generated Majorana masses a particular set of magnetized D9-branes
%engineers the Standard Model (or a GUT version of it), while the right-handed neutrinos $N_R^c$ arise
%as the bi-fundamental matter between a pair of $U(1)$ stacks with gauge groups $U(1)_b$ and $U(1)_c$.
% For $N_R^c$ transforming as, say, $(-1_b, 1_c)$, Majorana masses can be generated in the presence of precisely 2 %charged fermionic instanton zero modes of type $ \lambda_b$ and $\ov \lambda_c$ \cite{Blumenhagen:2006xt,Ibanez:2006da}.
%This guarantees that the coupling $\lambda_b N_R^c \ov\lambda_c$ in the instanton moduli action is
%allowed by gauge invariance. Integration over
%two copies of $\lambda$-modes then generates a mass term
%for $N_R^c$.

As an illustration we consider the generation of Majorana masses.
The GUT sector is located on a stack of $N_a=5$ $D9$-branes endowed
with line bundles $L_a$, while the right-handed neutrino $N_R^c$
transforms as $(-1_b, 1_c)$ of the gauge groups $U(1)_b$ and
$U(1)_c$ realised as single D9-branes carrying line bundles $L_b$
and $L_c$.  The zero mode constraints on the instanton for the
generation of Majorana masses translate into $h^i({\mathbf
P}^1,L_b^{\vee}(-1)|_{E_4}) = (2,0) = h^i({\mathbf
P}^1,L_c(-1)|_{E_4})$. For the subclass of ${\mathbf P}^1$s in
$\pi^*E_4$ with $d_9=0$, this  corresponds to
%Due to Bott's theorem  this is equivalent to
$\zeta_b \cdot E_4 = -2 = - \zeta_c \cdot E_4$. The actual presence
of the zero mode couplings  $c_{ijk} \lambda^i_b (N_R^c)^j \ov
\lambda^k_{c}$ in the moduli action is due to classical overlap
integrals and we have checked that they indeed exist. Further
details as well as the resulting family structure will be presented
in a forthcoming publication \cite{toappear}.

%Absence of any further charged zero
%modes %between the Standard Model sector and the instanton,
%between filler D5-branes and the instanton is guaranteed since the instanton wraps a non-horizontal curve.
%by the choice of the
%instanton as wrapping a ${\mathbb P}^1$ curve  in the del Pezzo surface
%$dP_9$ obtained as an elliptic fibration over $E_4$.
%Since our primary aim is to find global embeddings of the instanton sector, we content ourselves with
%the realisation of a GUT toy version of the Standard Model sector, namely to engineer it as a U(5)
%theory from $N_a = 5$ D9-branes with line bundle $L_a$.
%It turns out that the above constraints can easily be met in the present context.
In table \ref{model_Maj} we give a representative example of an
SU(5) model of the type described. All D5-brane tadpoles are
cancelled by including also stacks of $N_i$ unmagnetized D5-branes
on curves $\Gamma_i$ with total D5-brane charge $\sum N_i \gamma_i =
41 F + \sigma \cdot \pi^*(16 l -12 E_1)$, with $F$ the fibre class.
Furthermore, 12 unmagnetized D9-branes are required to cancel the
D9-brane tadpoles. One can check that the D-flatness conditions
allow for solutions inside the K\"ahler cone, e.g., for
$r_{\sigma}=1.00, r_l = 10.39, r_1=-7.00$ and $r_2=r_3=r_4=-1.00$.
\begin{table}[htb!]
\renewcommand{\arraystretch}{1.5}
\centering
\begin{tabular}{|c|c|c|}
\hline \hline
Bundle & N & $c_1(L)= q \sigma + \pi^*(\zeta)$    \\
\hline \hline $L_a$ & 5 & $\pi^*(-2 E_3)$ \\\hline $L_b$ & 1 & $2
\sigma + \pi^*(- 2l -2 E_1 + 3E_2 + 2E_3+ 2 E_4)$
\\\hline
$L_c$ & 1 & $-2 \sigma + \pi^*(2l - E_2 - 2 E_3 - 2 E_4)$    \\
\hline
%%%%%%%%%%%%%%%
\hline
\end{tabular}
\caption{\small A $ U(5) \times U(1) \times U(1)$ model with
Majorana masses.  } \label{model_Maj}
\end{table}
The spectrum can  be computed  from  the
formulae of \cite{Blumenhagen:2005zg,Blumenhagen:2005zh}. It
contains four chiral families of $\bf {\ov {10}}$ counted by $H^*(X,
(L_a^{\vee})^2)$ together with additional ${\bf 5}$ and ${\bf \ov
5}$ from the $a-b$ and $a-c$ sector as well as from the filler D5-
and D9-branes. Only part of them can be interpreted as matter ${\bf
5}$ and Higgs pairs once Yukawa couplings are taken into account.
%What is more of interest for us is the appearance of
Four chiral generations transforming as $(-1_b,1_c)$ in the spectrum
 %and counted by $\chi(X, L_b^{\vee} \otimes L_c)=4$.
 correspond to  right-handed neutrinos $N_R^c$. The zero mode structure of
the  $E1$ instanton ensures that a superpotential of the form $W =x
\,  M_s \exp({-  \frac{2 \pi \, \ell_s^4 }{\alpha_{GUT}}
\frac{{\mathcal Vol}}{{\widetilde f_a}} })\, N_R^c \,  N_R^c$ can be
generated, where $x$ is an ${\cal O}(1)$ factor from the exact
computation. The above value of the K\"ahler moduli corresponds to
${\widetilde f_a} = 9.56 \, \ell_s^6$ and an instanton volume
${\mathcal Vol}=-r_4=1$. Then for  $M_s \sim 10^{18}\, $GeV  and
$\alpha_{GUT}\sim 0.04$ the Majorana mass is  ${\cal O}(10^{11}) \,
$GeV. Thus, we demonstrated that in a global setup the exponential
suppression can indeed be engineered
at the required intermediate scale.  %While this is true for each single  instanton that wraps a ${\mathbb P}^1$ curve in the
%described class,
%, and satisfies the zero mode constraints,
The full answer involves summing up all instanton contributions
associated  with all suitable curves, but as alluded to already, depending on the concrete distribution of $D5$-branes required to cancel the tadpoles, it can
be arranged that most of the curves do not contribute due to extra $D5-E1$ zero modes. In favourable circumstances like the one given above it can even happen that only precisely one curve can contribute, in which case the associated superpotential term is guaranteed to be non-vanishing! 

Finally, we note that the examples presented in \cite{Cvetic:2007qj} can be further refined, and indeed Standard-like models with 3 chiral generations of matter exist in this framework while still realising Majorana masses in the instanton sector. These and further aspects will be discussed in \cite{toappear}.

\begin{acknowledgement}
We thank R. Blumenhagen and D. L\"ust for collaboration on \cite{Blumenhagen:2006xt,Blumenhagen:2007zk,Blumenhagen:2007bn} and \cite{Blumenhagen:2007zk}, respectively, as well as R. Donagi and T. Pantev for fruitful discussions. We are grateful to the organizers of the MCTP workshop  {\emph{The Physics and Mathematics of G2 Compactifications}}, String Phenomenology'07,  PASCOS'07,
BW'2007 and the Munich  {\emph{Workshop on recent developments in string effective actions and D-instantons}} for the invitation to present our ideas.
This research was supported in part by
DOE Grant EY-76-02-3071, and Fay R. and Eugene L. Langberg Chair funds.
\end{acknowledgement}

%\bibliographystyle{fdp}
%\bibliography{newproceedings.bib}

 %\nopagebreak
\renewcommand {\bibname} {\normalsize \sc References}
%\renewcommand {\refname} {\normalsize \sc References}
%\nopagebreak

{}

\end{document}